\documentclass[10pt,twocolumn,a4paper]{article}
\usepackage[margin=1.8cm,columnsep=0.8cm]{geometry}
\usepackage[T1]{fontenc}
\usepackage{amsmath,amssymb}
\usepackage{graphicx}
\usepackage{booktabs}
\usepackage{xcolor}
\usepackage[numbers,sort&compress]{natbib}
\usepackage[colorlinks=true,allcolors=blue!50!black]{hyperref}

\newcommand{\Pme}{P_{\mu e}}
\newcommand{\dPme}{\Delta P_{\mu e}}
\newcommand{\Dmsq}[1]{\Delta m^2_{#1}}
\newcommand{\Rearth}{R_\oplus}
\newcommand{\dCP}{\delta_{\rm CP}}
\newcommand{\numu}{\nu_\mu}
\newcommand{\nue}{\nu_e}

\title{Earthquake Forecast via Neutrino Tomography at the DUNE Baseline}
\author{%
Bipin Singh Koranga\textsuperscript{1},\;
Samrat Vashisht\textsuperscript{1},\;
Vivek Kumar Nautiyal\textsuperscript{2}\\[4pt]
{\small\textsuperscript{1}Department of Physics, Kirori Mal College, University of Delhi, Delhi--110007, India}\\
{\small\textsuperscript{2}Department of Physics, Chaudhary Charan Singh University, Meerut, India}}
\date{}

\begin{document}
\makeatletter
\twocolumn[\begin{@twocolumnfalse}
\maketitle
\begin{abstract}
We investigate whether the matter effect on accelerator neutrino oscillations over the 1300~km Fermilab--SURF baseline of DUNE could register an anomalous electron-density accumulation in a seismic fault zone lying on the beam path, following the reactor-antineutrino proposal of Wang, Chen and Li. The fault zone is modelled as a uniform slab of width $w$ and effective density $\eta\rho_0$ embedded in normal crust, and the full three-flavour appearance and disappearance probabilities are computed with exact slab-by-slab evolution. For $\eta=3$ and $w=20$--$100$~km we find shifts of $\dPme\simeq0.1$--$0.55\%$ (absolute) for neutrinos and $-0.08$ to $-0.37\%$ for antineutrinos at $E_\nu=2.5$~GeV, i.e.\ relative changes of $1.4$--$6.8\%$ ($\nu$) and $3.5$--$17\%$ ($\bar\nu$) in the appearance probability, and up to $0.6\%$ in $P_{\mu\mu}$. The shift is nearly independent of $\dCP$, so a time-differential comparison cancels the CP-phase degeneracy. A purely statistical $3\sigma$ detection requires $\mathcal{O}(10^2$--$10^4)$ appearance events within the anomaly window, which is far above what a week-long precursor window can provide at DUNE. Detection of a long-lived density anomaly is marginal; short-term forecasting is not feasible with the nominal beam. The fault must also lie on the beam chord at an appropriate depth. We regard the result as a theoretical sensitivity study rather than a forecasting scheme.
\end{abstract}

\noindent\textbf{Keywords:} neutrino oscillations; matter effects; neutrino tomography; DUNE; long-baseline experiments; earthquake precursors; CP violation.\\
\textbf{PACS:} 14.60.Pq (neutrino mass and mixing); 14.60.Lm (ordinary neutrinos); 91.30.Px (earthquake prediction and precursors); 91.35.$-$x (structure of the Earth's interior).
\bigskip
\end{@twocolumnfalse}]
\makeatother
\nocite{DeRujula:1983,Wolfenstein:1978,MikheevSmirnov:1985,Barger:1980,OhlssonWinter:2002,OhlssonWinter:2003,Winter:2006,Lindner:2003,GonzalezGarcia:2008,Rott:2015,Donini:2019,Wang:2010,Pulinets:2004,Dobrovolsky:1979,Geller:1997,PREM:1981,DUNE:CDR,DUNE:TDR,DUNE:physics,HyperK:DR,Cervera:2000,Gandhi:2006,Nunokawa:2008,OhlssonSnellman:2000,Kimura:2002,Freund:2001,Akhmedov:2004,MinakataNunokawa:2001,IoannisianSmirnov:2004,NuFit6:2024,PDG:2024,Cowan:2011,Koranga:code}

\section{Introduction}
The idea that neutrinos can be used to image the interior of the Earth is almost as old as the neutrino itself. Early proposals relied on the attenuation of high-energy neutrino fluxes crossing the planet~\cite{DeRujula:1983}. After the Mikheyev--Smirnov--Wolfenstein (MSW) mechanism~\cite{Wolfenstein:1978,MikheevSmirnov:1985} established that coherent forward scattering on electrons modifies flavour conversion in matter~\cite{Barger:1980}, attention turned to oscillation tomography, in which the energy and baseline dependence of the oscillation probabilities encodes the electron density along the path. This approach has been developed for long-baseline beams, for example to constrain the line-averaged crustal density or to search for localised density anomalies such as oil deposits~\cite{OhlssonWinter:2002,OhlssonWinter:2003,Winter:2006}, and for atmospheric neutrinos and supernova neutrinos, in order to probe the mantle and core~\cite{Lindner:2003,GonzalezGarcia:2008,Rott:2015}. The prospects of the field have recently been reviewed in Ref.~\cite{Donini:2019}.

Wang, Chen and Li~\cite{Wang:2010} proposed a geophysical application of this idea: the matter effect on (anti)neutrino oscillations could serve as a tomographic probe of the electron density in a seismic fault zone. Their motivation was that a number of precursory phenomena are reported to precede strong earthquakes, and that some of them could involve an anomalous increase of the electron density in the fault region~\cite{Pulinets:2004}. Using reactor antineutrinos with $E_\nu=3.6$~MeV over a baseline of order $10^3$~km, they found a survival-probability variation of up to $3\%$ for an anomalous density increase $\eta=3$, and concluded that accelerator neutrinos of $E_\nu=7$~GeV give an unobservable ($0.01\%$) effect for a 6000~km baseline in their two-flavour treatment. It should be said at the outset that the reliability of earthquake precursors, and the very possibility of short-term earthquake prediction, remain controversial in the geophysics community~\cite{Geller:1997}. The present work does not address that question; it asks only what a neutrino experiment could measure if a density anomaly of the assumed type existed.

Since Ref.~\cite{Wang:2010} the long-baseline programme has matured. The Deep Underground Neutrino Experiment (DUNE)~\cite{DUNE:CDR,DUNE:TDR,DUNE:physics} will send a wide-band $\numu$ beam over $L=1300$~km, with the first oscillation maximum near $E_\nu\simeq2.5$~GeV. At this energy the matter potential is comparable to the atmospheric splitting, $2E_\nu V/\Dmsq{31}\sim0.2$ in normal crust, so the baseline is strongly sensitive to the integrated electron density, and this is the reason DUNE is sensitive to the mass ordering~\cite{Cervera:2000,Gandhi:2006,Nunokawa:2008}. This work asks the natural question: if the same physics is applied with the DUNE beam and a full three-flavour treatment, what is the shift caused by a fault zone on the baseline, and is it observable?

The main new elements relative to Ref.~\cite{Wang:2010} are:
\begin{enumerate}
\item[(i)] an exact three-flavour, slab-by-slab treatment at a 1300~km accelerator baseline, valid in the non-perturbative regime inside the anomaly;
\item[(ii)] a quantitative demonstration that the fault-induced shift is nearly $\dCP$-independent, so that a time-differential analysis removes the CP-phase degeneracy; and
\item[(iii)] an explicit statistical benchmark (event counts for a $3\sigma$ detection) that identifies which features of an anomaly are experimentally relevant.
\end{enumerate}
The paper is organised as follows. Section~\ref{sec:model} describes the fault-zone model and the beam geometry. Section~\ref{sec:formalism} gives the three-flavour slab formalism (including the explicit evolution equations and numerical checks) and the oscillation parameters. Section~\ref{sec:results} presents the results, Sec.~\ref{sec:stat} the statistical requirement for detection, and Sec.~\ref{sec:caveats} the caveats. Section~\ref{sec:conclusion} concludes.

\section{Fault-zone model and baseline geometry}\label{sec:model}
We keep the fault-zone model of Ref.~\cite{Wang:2010}. Before a strong earthquake the electron density in the fault zone is taken to rise from the normal crustal value $n_e^0=\rho_0/(2m_N)$ to
\begin{equation}
n_e^f=\eta\,n_e^0,\qquad \rho_f'=\eta\rho_0,
\label{eq:eta}
\end{equation}
with $Y_e\simeq0.5$ and $\rho_0=2.8~\mathrm{g/cm^3}$, a value representative of continental crust and lying between the upper- and lower-crust densities of the Preliminary Reference Earth Model~\cite{PREM:1981}. The width of the fault zone is related to the surface-wave magnitude $M_S$ of the impending event by the empirical formula~\cite{Pulinets:2004,Dobrovolsky:1979}
\begin{equation}
w=2\times10^{\,0.414M_S-1.696}~\mathrm{km},
\label{eq:width}
\end{equation}
which gives $w\approx80$~km for $M_S=8$. As in Ref.~\cite{Wang:2010}, $\rho_f'$ is an effective density that encodes the electron content only; the true mass density $\rho_f$ may be smaller than $\eta\rho_0$.

\begin{figure*}[!t]
\centering
\includegraphics[width=0.7\textwidth]{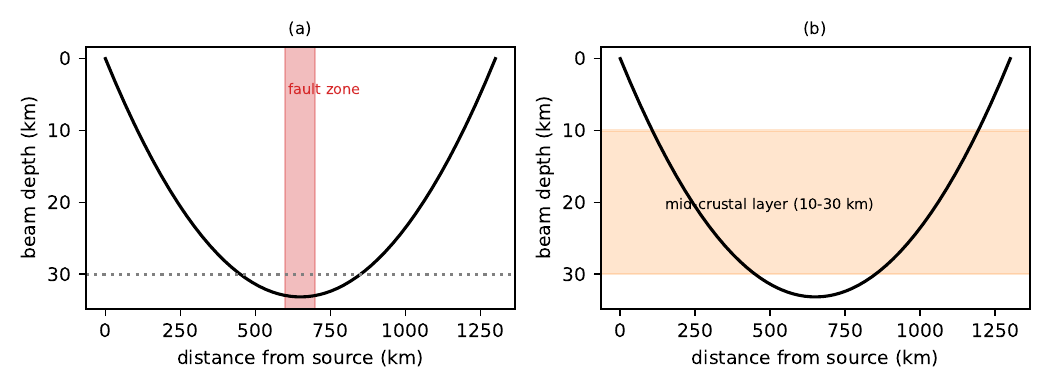}
\caption{(a) Depth of the 1300~km chord below the surface, with a fault zone of width 100~km centred at the midpoint. (b) Chord depth compared with a 10--30~km mid-crustal layer. The chord depth follows $d(x)\simeq x(L-x)/2\Rearth$, reaching about 33~km at $x=650$~km; the shaded regions mark the fault zone (a) and the mid-crustal layer (b), so only the parts of the chord inside the shading are sensitive to the anomaly.}
\label{fig:geom}
\end{figure*}

The beam travels along a chord of length $L=1300$~km, so its depth at a distance $x$ from the source is
\begin{equation}
d(x)\simeq\frac{x(L-x)}{2\Rearth},
\label{eq:depth}
\end{equation}
which reaches $\approx33$~km at the midpoint and equals about 9, 14 and 17~km at $x=100$, 150 and 200~km (Fig.~\ref{fig:geom}). A mid-crustal fault zone of the kind considered in Ref.~\cite{Wang:2010} (a 10~km deep cuboid in a 30~km crust) is therefore intersected by the beam only within a few hundred km of either end of the chord; near the midpoint the beam is in the lower crust or at the Moho. We treat the zone as a slab of density $\eta\rho_0$ and thickness $w$ along the beam, centred at $x_0$, with normal crust ($\rho_0$) elsewhere (Fig.~\ref{fig:geom}). The three slabs are called I, II, III as in Ref.~\cite{Wang:2010}.

\section{Three-flavour slab formalism}\label{sec:formalism}
\subsection{Hamiltonian and observable}
In the flavour basis the Hamiltonian in a slab of density $\rho$ is
\begin{align}
H&=\frac{1}{2E_\nu}\,U\,\mathrm{diag}(0,\Dmsq{21},\Dmsq{31})\,U^\dagger\nonumber\\
&\quad+\mathrm{diag}(V,0,0),\label{eq:H}\\
V&=\sqrt2\,G_F\,n_e\nonumber\\
&\simeq7.63\times10^{-14}\,Y_e\,\frac{\rho}{\mathrm{g\,cm^{-3}}}~\mathrm{eV},\label{eq:V}
\end{align}
where $U=U(\theta_{23},\theta_{13},\theta_{12},\dCP)$ is the standard parametrisation of the lepton mixing matrix~\cite{PDG:2024}. For antineutrinos $V\to-V$ and $\dCP\to-\dCP$. The amplitude through the three slabs is
\begin{equation}
\begin{split}S&=e^{-iH_{III}L_{III}}\,e^{-iH_{II}w}\,e^{-iH_{I}L_{I}},\\ P_{\alpha\beta}&=|S_{\beta\alpha}|^2,\end{split}
\label{eq:S}
\end{equation}
which is the three-flavour generalisation of the $R$-matrix product used in Ref.~\cite{Wang:2010} (their Eq.~(10)), evaluated here by exact matrix exponentiation rather than the two-flavour reduction. Exact treatments of three-flavour propagation in constant matter are well established~\cite{OhlssonSnellman:2000,Kimura:2002}, and compact analytic approximations are available~\cite{Freund:2001,Akhmedov:2004,MinakataNunokawa:2001}, but the latter assume a small expansion parameter that is not satisfied inside a dense anomaly (see Sec.~\ref{sec:caveats}). Unlike the reactor case, the $\theta_{13}\to0$ reduction of Ref.~\cite{Wang:2010} is not valid at DUNE, since $\numu\to\nue$ appearance is itself driven by $\theta_{13}$.

We use $\sin^2\theta_{12}=0.307$, $\sin^2\theta_{13}=0.0220$, $\sin^2\theta_{23}=0.546$, $\dCP=1.23\pi$, $\Dmsq{21}=7.49\times10^{-5}~\mathrm{eV^2}$ and $\Dmsq{31}=2.513\times10^{-3}~\mathrm{eV^2}$ (normal ordering), representative of global fits~\cite{NuFit6:2024,PDG:2024}. The observable is
\begin{equation}
\Delta P_{\alpha\beta}=P_{\alpha\beta}(\text{fault})-P_{\alpha\beta}(\text{no fault}),
\label{eq:dP}
\end{equation}
the analogue of the quantity plotted in Figs.~2--4 of Ref.~\cite{Wang:2010}.

\subsection{Explicit slab-by-slab evolution and numerical procedure}
For a fault zone of width $w$ centred at $x_0$ on a chord of length $L$, the three path segments are
\begin{equation}
\begin{split}L_I&=x_0-\frac w2,\qquad L_{II}=w,\\ L_{III}&=L-x_0-\frac w2,\end{split}
\label{eq:segments}
\end{equation}
with densities $\rho_I=\rho_{III}=\rho_0$ and $\rho_{II}=\eta\rho_0$, so that $L_I+L_{II}+L_{III}=L$. (This assumes the zone lies entirely on the chord, $w/2\le x_0\le L-w/2$; in the accompanying code a zone that overruns either end is truncated at the chord end, so the total path length is always $L$.) Each slab Hamiltonian $H_i\equiv H(E_\nu,\rho_i)$ of Eq.~\eqref{eq:H} is Hermitian and is diagonalised as
\begin{equation}
\begin{split}H_i&=\sum_{k=1}^3\lambda_k^{(i)}|\nu_k^{(i)}\rangle\langle\nu_k^{(i)}|,\\
e^{-iH_iL_i}&=\sum_{k=1}^3e^{-i\lambda_k^{(i)}L_i}|\nu_k^{(i)}\rangle\langle\nu_k^{(i)}|,\end{split}
\label{eq:diag}
\end{equation}
where $\lambda_k^{(i)}$ are the roots of the characteristic cubic of $H_i$ (the matter eigenvalues, obtainable in closed form~\cite{Kimura:2002}). The slab matrices are then multiplied in the order of Eq.~\eqref{eq:S},
\begin{equation}
S=U_{III}\,U_{II}\,U_{I},\qquad U_i\equiv e^{-iH_iL_i},
\label{eq:Smatrix}
\end{equation}
and the appearance and disappearance probabilities are $\Pme=|S_{e\mu}|^2$ and $P_{\mu\mu}=|S_{\mu\mu}|^2$. The matter potential is $V\to-V$, $\dCP\to-\dCP$ for antineutrinos. Distances are converted to natural units with $1~\mathrm{km}=5.068\times10^{9}~\mathrm{eV^{-1}}$, and $E_\nu$ is in eV. The transition matrix is unitary for every slab, hence for the product; no averaging over the spectrum or over production and detection is applied (the probabilities are for a monochromatic beam). In an independent re-evaluation of the calculation we found that $S^\dagger S=\mathbb{1}$ holds to better than $10^{-15}$ in double precision, and that the values of $\dPme$ and the no-fault probabilities quoted in Sec.~\ref{sec:results} are reproduced; the code is available in Ref.~\cite{Koranga:code}. Table~\ref{tab:matter} summarises the matter-modified quantities used in the discussion of the anomaly and in Sec.~\ref{sec:caveats}. The magnitude of the matter parameter $2E_\nu V/\Dmsq{31}$ explains why the exact (non-perturbative) evolution is required inside slab~II.

\begin{table}[t]
\centering\small\setlength{\tabcolsep}{3pt}
\caption{Matter-modified quantities at $E_\nu=2.5$~GeV: splitting between the heaviest and lightest matter eigenstates, and electron fraction of the heaviest eigenstate (vacuum value $\sin^2\theta_{13}=0.022$); the middle state in slab~II has $|U^M_{e2}|^2=0.864$.}
\label{tab:matter}
\begin{tabular}{lccc}
\toprule
Slab & $\rho$ (g/cm$^3$) & $\Delta m^2_{\rm eff}$ ($10^{-3}$ eV$^2$) & $|U^M_{e3}|^2$\\
\midrule
I, III & 2.8 & 2.48 & 0.035\\
II ($\eta=3$) & 8.4 & 2.55 & 0.136\\
\bottomrule
\end{tabular}
\end{table}

\section{Results}\label{sec:results}\begin{figure*}[!t]
\centering
\includegraphics[width=0.7\textwidth]{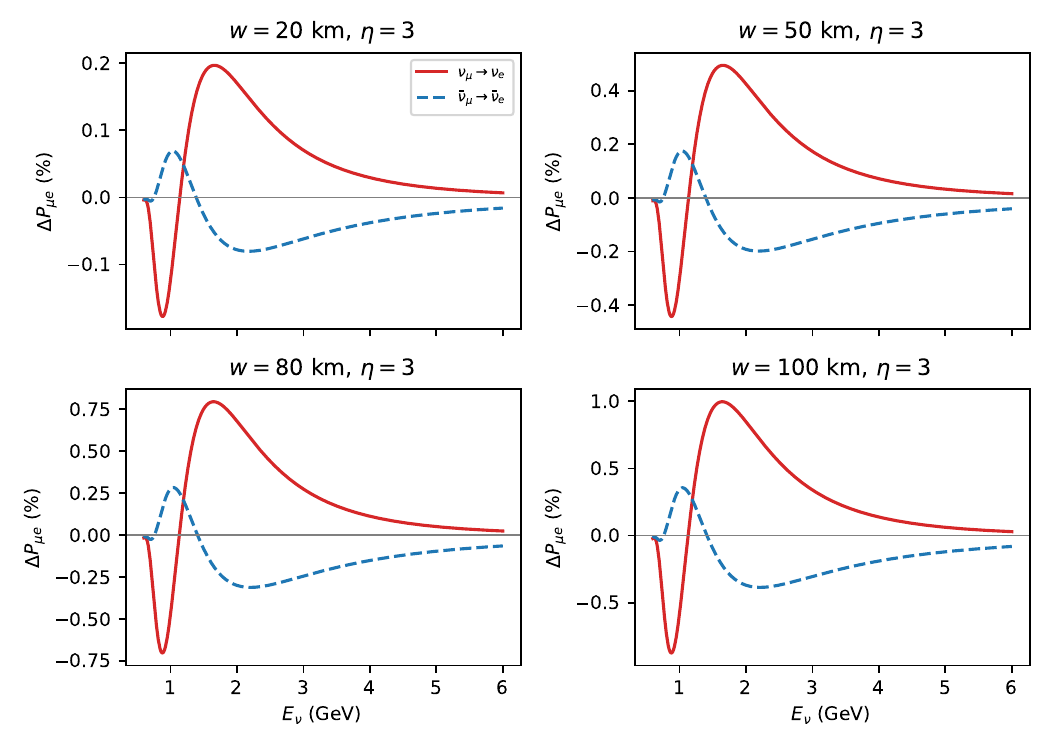}
\caption{Fault-induced change in $\Pme$ (percent, absolute) for $\eta=3$ and $w=20,50,80,100$~km, for neutrinos (solid) and antineutrinos (dashed). The fault zone is centred at the midpoint of the 1300~km baseline. The neutrino shift changes sign near $E_\nu\simeq1.1$~GeV and the antineutrino shift near 0.8 and 1.4~GeV; the two shifts are opposite in sign for $E_\nu\gtrsim1.4$~GeV, and the shift grows approximately linearly with $w$.}
\label{fig:E}
\end{figure*}

\paragraph{Energy dependence.} Figure~\ref{fig:E} shows $\dPme$ versus $E_\nu$ for $\eta=3$ and $w=20,50,80,100$~km with the fault at the midpoint. The shift grows roughly linearly with $w$. At $E_\nu=2.5$~GeV the neutrino shifts are $+0.11,+0.28,+0.44,+0.55\%$ (absolute), i.e.\ relative to $\Pme=0.081$ of $1.4,3.4,5.5,6.8\%$; for antineutrinos ($P_{\bar\mu\bar e}=0.022$) they are $-0.08,-0.19,-0.30,-0.37\%$, i.e.\ $-3.5,-8.6,-13.6,-16.9\%$. The opposite signs for $\nu$ and $\bar\nu$ at these energies follow from the sign of $V$ and provide a useful consistency check; the two shifts have opposite sign for $E_\nu\gtrsim1.4$~GeV and, for $w=100$~km, also below $1.1$~GeV, but share the same (positive) sign in the narrow window $1.1\lesssim E_\nu\lesssim1.4$~GeV between their zero crossings. The neutrino shift changes sign near $E_\nu\simeq1.1$~GeV (e.g.\ $-0.59\%$ for $\nu$ at $E_\nu=1$~GeV and $w=100$~km), which gives a characteristic spectral signature.

\begin{figure*}[!t]
\centering
\includegraphics[width=0.7\textwidth]{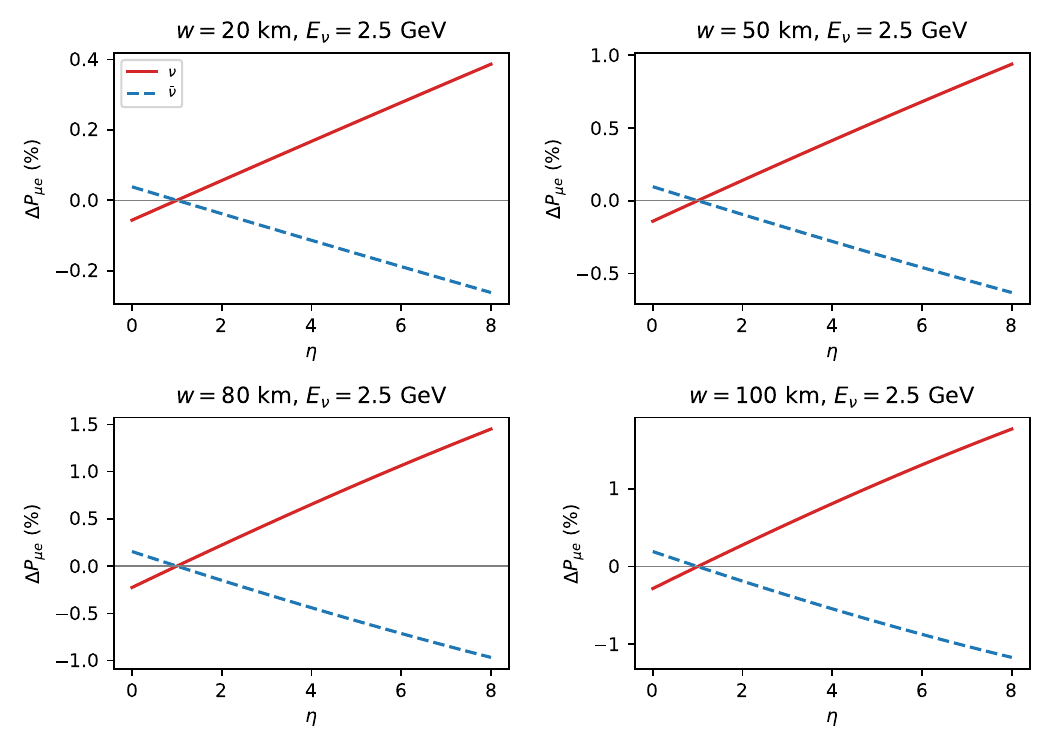}
\caption{$\dPme$ versus $\eta$ at $E_\nu=2.5$~GeV for four widths. Solid (dashed) curves are neutrinos (antineutrinos); the shift vanishes at $\eta=1$ (no anomaly) and changes sign for $\eta<1$.}
\label{fig:eta}
\end{figure*}

\paragraph{Dependence on $\eta$.} Figure~\ref{fig:eta} shows the dependence on $\eta$ at $E_\nu=2.5$~GeV. As in Ref.~\cite{Wang:2010} the effect vanishes at $\eta=1$ and increases with $\eta$; for $\eta<1$ the sign flips because the zone is then less dense than the surrounding crust.

\begin{figure*}[!t]
\centering
\includegraphics[width=0.7\textwidth]{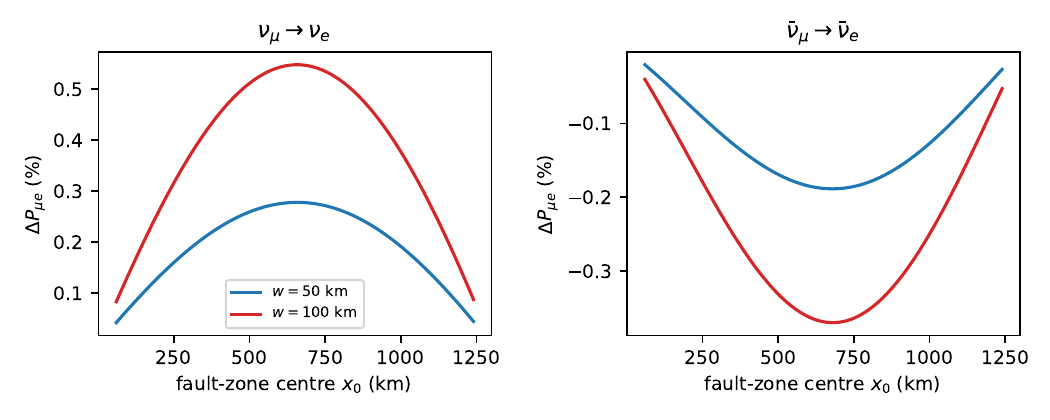}
\caption{$\dPme$ versus the fault-zone centre $x_0$ for $E_\nu=2.5$~GeV, $\eta=3$. Left: neutrinos; right: antineutrinos; blue and red curves correspond to $w=50$ and $100$~km. The shift is largest for a zone at the midpoint and falls by a factor of about four towards the ends of the chord, so that $x_0$ is strongly entangled with $w$ and $\eta$ when only a single baseline is available.}
\label{fig:x0}
\end{figure*}

\paragraph{Position of the zone.} Figure~\ref{fig:x0} shows how $\dPme$ depends on the zone centre $x_0$. For $w=100$~km the neutrino shift is $0.14\%$ at $x_0=100$~km, $0.37\%$ at $x_0=300$~km and $0.55\%$ at the midpoint, i.e.\ it varies by a factor of about four across the baseline and is largest for a central zone, reflecting the position-dependent phase accumulated before and after the slab. To first order the \emph{size} of the effect is controlled by the length-averaged density, $\bar\rho=\rho_0[1+(\eta-1)w/L]$, which gives the correct sign and order of magnitude (e.g.\ $+0.18\%$ against the exact $+0.28\%$ for $w=50$~km, and $+0.36\%$ against $+0.55\%$ for $w=100$~km) but, being position-independent, cannot reproduce the $x_0$ dependence. The difference reflects the fact that the density perturbation is localised rather than averaged. Practical consequence: the signal is large only for wide, strongly anomalous zones, and because the amplitude is degenerate in $w$, $\eta$ and $x_0$, a tomographic reading of the location from a single baseline and a single energy is weak.

\begin{figure}[!t]
\centering
\includegraphics[width=0.85\columnwidth]{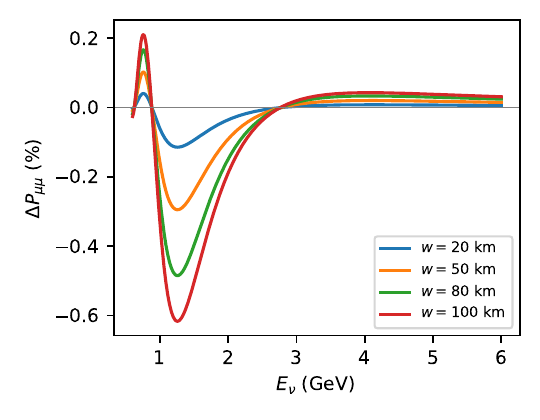}
\caption{$\Delta P_{\mu\mu}$ versus $E_\nu$ for $\eta=3$. Curves correspond to $w=20,50,80,100$~km; the disappearance shift is at most $0.62\%$ in magnitude, near $E_\nu\simeq1.2$~GeV.}
\label{fig:mumu}
\end{figure}

\paragraph{Disappearance channel.} The $\numu\to\numu$ channel shifts by at most $0.62\%$ (absolute) at $w=100$~km, $\eta=3$ (Fig.~\ref{fig:mumu}), small compared with the unit-level probability but statistically favourable because the $\numu$ event sample is two orders of magnitude larger than the $\nue$ sample.

\begin{figure*}[!t]
\centering
\includegraphics[width=0.7\textwidth]{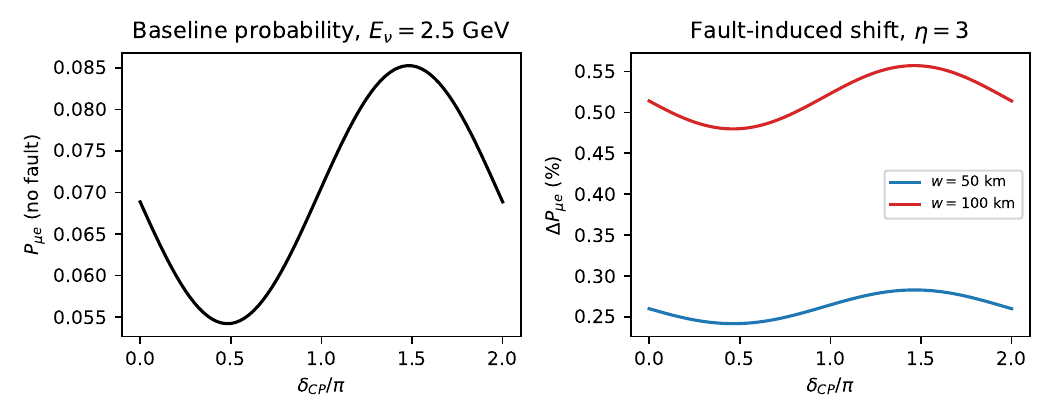}
\caption{Left: no-fault $\Pme$ versus $\dCP$. Right: fault-induced shift for $\eta=3$, $E_\nu=2.5$~GeV. While the baseline probability varies by tens of percent across $\dCP$, the fault-induced shift varies by less than $0.1$ percentage points for both $w=50$~km (blue) and $w=100$~km (red).}
\label{fig:cp}
\end{figure*}

\paragraph{CP-phase degeneracy.} A static shift in $\Pme$ is degenerate with $\dCP$ and with the average density~\cite{Nunokawa:2008,MinakataNunokawa:2001}: the baseline probability itself changes from $0.054$ to $0.085$ at 2.5~GeV across $\dCP\in[0.5\pi,1.5\pi]$, i.e.\ by tens of percent. In contrast, the fault-induced shift is almost independent of $\dCP$ ($0.48$--$0.56\%$ for $w=100$~km and $0.24$--$0.28\%$ for $w=50$~km; Fig.~\ref{fig:cp}). A time-differential comparison of the event spectrum before and after (or in successive periods, as Ref.~\cite{Wang:2010} suggests) therefore cancels $\dCP$, $\theta_{23}$ and the average-density uncertainty~\cite{OhlssonWinter:2003}, and leaves only detector, flux and calibration stability as the main systematic limits.

\section{Statistical requirement}\label{sec:stat}
For a counting experiment dominated by statistical error, a relative change $\delta=\Delta P/P$ in an event sample of $N$ events is detected at $z\sigma$ when $N\gtrsim z^2/\delta^2$ (see Ref.~\cite{Cowan:2011} for the treatment of the general case). For $z=3$ and $E_\nu=2.5$~GeV the required number of $\nue$ appearance events in each time bin is given in Table~\ref{tab:stat}; for $|\delta|\gtrsim0.3$ (last row) this linearised estimate is only indicative.

\begin{table}[t]
\centering\small\setlength{\tabcolsep}{3pt}
\caption{Number of $\nu_e$ ($\bar\nu_e$) appearance events $N$ needed for a $3\sigma$ statistical detection at $E_\nu=2.5$~GeV.}
\label{tab:stat}
\begin{tabular}{rcccccc}
\toprule
$w$ (km) & $\eta$ & $|\delta_\nu|$ & $N_\nu$ & $|\delta_{\bar\nu}|$ & $N_{\bar\nu}$\\
\midrule
 20 & 3  & 0.014 & $4.7\times10^4$ & 0.035 & $7.5\times10^3$\\
 50 & 3  & 0.034 & $7.6\times10^3$ & 0.086 & $1.2\times10^3$\\
100 & 3  & 0.068 & $2.0\times10^3$ & 0.169 & $3.2\times10^2$\\
100 & 10 & 0.27  & $1.2\times10^2$ & 0.65  & 21\\
\bottomrule
\end{tabular}
\end{table}

The nominal DUNE exposure is expected to yield an appearance sample of order $10^3$--$10^4$ events accumulated over several years~\cite{DUNE:physics} (to be replaced by a full simulation with the DUNE flux, cross sections and detector response). Thus: (i) a slowly developing anomaly sustained over years, comparing yearly samples, could marginally be probed only for wide, strongly anomalous zones ($w\sim100$~km, $\eta\gtrsim3$) in antineutrino mode; (ii) a precursor lasting one week ($\sim1/50$ of a year) leaves only a few events and is not accessible at the level required. This conclusion is more pessimistic than the optimistic statement in Ref.~\cite{Wang:2010} that an enhanced detection efficiency allows medium-term forecasts, because the DUNE beam is not tunable in the way an idealised flux is, and the zone must sit on the fixed beam chord. Larger far detectors, such as that proposed for Hyper-Kamiokande~\cite{HyperK:DR}, would improve the statistics, but their baselines are likewise fixed by the source location.

\section{Discussion and caveats}\label{sec:caveats}
\begin{enumerate}
\item \emph{Geology.} The fixed chord from Fermilab to the Sanford Underground Research Facility crosses the stable continental interior. We do not claim that an active fault zone of the assumed width lies on it, so the study is a sensitivity estimate. A dedicated experiment would place the baseline across a known seismic zone.
\item \emph{The value of $\eta$.} As emphasised in Ref.~\cite{Wang:2010}, $\eta$ is highly uncertain and controversial. Values above 1 produced by strain, shear, dehydration or an anomalous electric field have not been independently established~\cite{Geller:1997}. We adopt $\eta=3$ for comparison only.
\item \emph{Effective density.} The electron-density anomaly need not correspond to a change of the true mass density. The neutrino signal fixes the electron column density $\propto\rho_f'$, not $\rho_f$.
\item \emph{Backgrounds and systematics.} Uncertainty in the line-averaged crustal density (a few per cent)~\cite{OhlssonWinter:2003}, flux normalisation, near-to-far extrapolation and neutral-current and $\nu_\tau$ backgrounds are ignored. They affect a time-differential comparison only through drifts, which are not modelled here.
\item \emph{Perturbative regime.} The expansion parameter $2E_\nu V/\Dmsq{31}$ is $\approx0.21$ in normal crust and $\approx0.64$ in the anomaly zone, so the Ioannisian--Smirnov-type expansion of Refs.~\cite{Wang:2010,IoannisianSmirnov:2004} is not accurate here and exact slab evolution is needed.
\end{enumerate}

\section{Conclusion}\label{sec:conclusion}
Extending the neutrino-tomography proposal of Ref.~\cite{Wang:2010} to the DUNE baseline with a full three-flavour calculation, we find that a fault zone with $\eta=3$ and $w=20$--$100$~km changes $\Pme$ by $0.1$--$0.55\%$ (absolute) at 2.5~GeV, with opposite sign for antineutrinos and a nearly $\dCP$-independent magnitude. This makes a time-differential measurement conceptually clean. However, the required event counts exceed what the nominal DUNE beam can accumulate on short precursor timescales, and the zone must lie on the beam chord. DUNE is therefore not a practical earthquake-forecasting instrument; the study instead gives a benchmark for the sensitivity of future long-baseline experiments to localised matter-density anomalies, and shows which of the proposed features (large $w$, antineutrino mode, time-differential analysis) matter most.

\section*{Data availability}
All numerical results are obtained from the exact slab evolution of Sec.~\ref{sec:formalism}B with the parameters listed in Sec.~\ref{sec:formalism}A; no experimental data were used. The Python code (\texttt{slab.py}, \texttt{make\_figures.py}, and tests) that reproduces Figs.~\ref{fig:geom}--\ref{fig:cp} and Tables~\ref{tab:matter} and~\ref{tab:stat} is archived in Ref.~\cite{Koranga:code}, together with the generated figures and a README with the conventions and usage.

\bibliographystyle{unsrtnat}
\bibliography{refs}
\end{document}